\documentclass[amsmath,amssymb,aps,prl,reprint,superscriptaddress,footinbib]{revtex4-1}

\usepackage{tensor}
\usepackage{graphicx}
\usepackage{amsmath, amssymb, amsfonts, mathtools}
\usepackage{ulem}
\usepackage{float}
\usepackage{color}
\usepackage{natbib}

\newcommand\apjl{ApJ}

\newcommand\mnras{MNRAS}

\renewcommand{\vec}[1]{\boldsymbol{#1}} % Standard \vec{} gives arrow above

\begin{document}

\title{Origin of Pulsar Radio Emission}

\author{Alexander Philippov}
\email{sphilippov@flatironinstitute.org}
\affiliation{Center for Computational Astrophysics, Flatiron Institute, 162 Fifth Avenue, New York, NY 10010, USA}

\author{Andrey Timokhin}
\affiliation{Janusz Gil Institute of Astronomy, University of Zielona G\'{o}ra, ul. Szafrana 2, 65–516 Zielona G\'{o}ra, Poland}

\author{Anatoly Spitkovsky}
\affiliation{Department of Astrophysical Sciences, Princeton University, 4 Ivy Lane, Princeton, NJ 08544, USA}

\begin{abstract}
Since pulsars were discovered as emitters of bright coherent radio emission more than half a century ago, the cause of the emission has remained a mystery. In this Letter we demonstrate that coherent radiation can be directly generated in non-stationary pair plasma discharges which are responsible for filling the pulsar magnetosphere with plasma. By means of large-scale two-dimensional kinetic plasma simulations, we show that if pair creation is non-uniform across magnetic field lines, the screening of electric field by freshly produced pair plasma is accompanied by the emission of waves which are electromagnetic in nature. 
Using localized simulations of the screening process, we identify these waves as superluminal ordinary (O) modes, which should freely escape from the magnetosphere as the plasma density drops along the wave path. The spectrum of the waves is broadband and the frequency range is comparable to that of observed pulsar radio emission. 
\end{abstract}

\maketitle

Pulsars are rapidly rotating highly magnetized neutrons stars (NS), most of them are sources of coherent radio emission. It is universally accepted that active pulsars generate dense electron-positron plasma, which fills their magnetospheres. The main channel of pair creation is believed to be the absorption of $\gamma$-rays in super-strong magnetic fields \cite{Sturrock71}. Pulsar radio emission has remained an enigmatic phenomenon since its discovery. Early analytical theories advocated for plasma instabilities that could be excited in the uniform pair plasma outflow above the pair formation front. The most popular ideas invoked two-stream instabilities and conversion of excited plasma waves into escaping electromagnetic radiation, or the emission of coherent curvature radiation by charge bunches \cite{Ruderman75}. The growth of two-stream instability is severely reduced by relativistic streaming of pair plasma (unless the overlap of distinct plasma clouds is invoked \cite{Usov87}), and it is unclear whether efficient wave conversion even happens. Theories involving charged bunches face severe difficulties of forming long-lived bunches in the first place \cite{Melrose17}. Moreover, the last decade of kinetic plasma simulations of discharges in the pulsar magnetosphere revealed their essentially time-dependent nature, which questions any steady-state theory. 

It has been demonstrated by direct numerical simulations \cite{Timokhin10,Timokhin13} that electron-positron pair creation in pulsar polar caps -- regions near NS magnetic poles -- always proceeds via intermittent discharges. Each discharge starts with the formation of a gap -- a charge-starved region with a strong electric field -- where particles are accelerated to high energies until they start emitting pair-producing photons. Newly born pairs screen the accelerating electric field, thus preventing further particle acceleration. When the pair plasma leaves the polar cap, a new discharge begins. Screening of the accelerating field involves large amplitude fluctuations of the electric field and collective plasma motions. It is reasonable to expect that such screening events can produce coherent electromagnetic radiation directly {\cite{Beloborodov08,Timokhin10,Timokhin13}}. 

The problem with such a straightforward mechanism is that observable electromagnetic waves should propagate in the general direction of the background magnetic field, which requires the electric field of the wave to be transverse to the background magnetic field. Excitation of such waves requires transverse charge or current fluctuations; however, in the super strong magnetic field near polar caps charged particles can move only along the magnetic field lines.  Moreover, in order to be observable, these waves have to propagate through the magnetosphere filled with dense pair plasma without substantial damping. 

In this Letter, we argue that the inevitable non-uniformity of pair formation across magnetic field lines results in a fluctuating component of the electric field perpendicular to the background magnetic field and in the excitation of transverse waves. By means of first-principles kinetic plasma simulations we investigate which modes are produced in the process of a non-stationary pair plasma discharge near the NS surface. We demonstrate that the non-uniformity of pair formation across magnetic field lines leads to direct radiation of superluminal ordinary electromagnetic waves, which do not suffer from Landau damping and should be able to freely escape from the magnetosphere.

\begin{figure*}
    %\label{fig:my_label}
        \includegraphics[width=1.\textwidth, trim = 10.0mm 8.0mm 10.0mm 4.0mm]{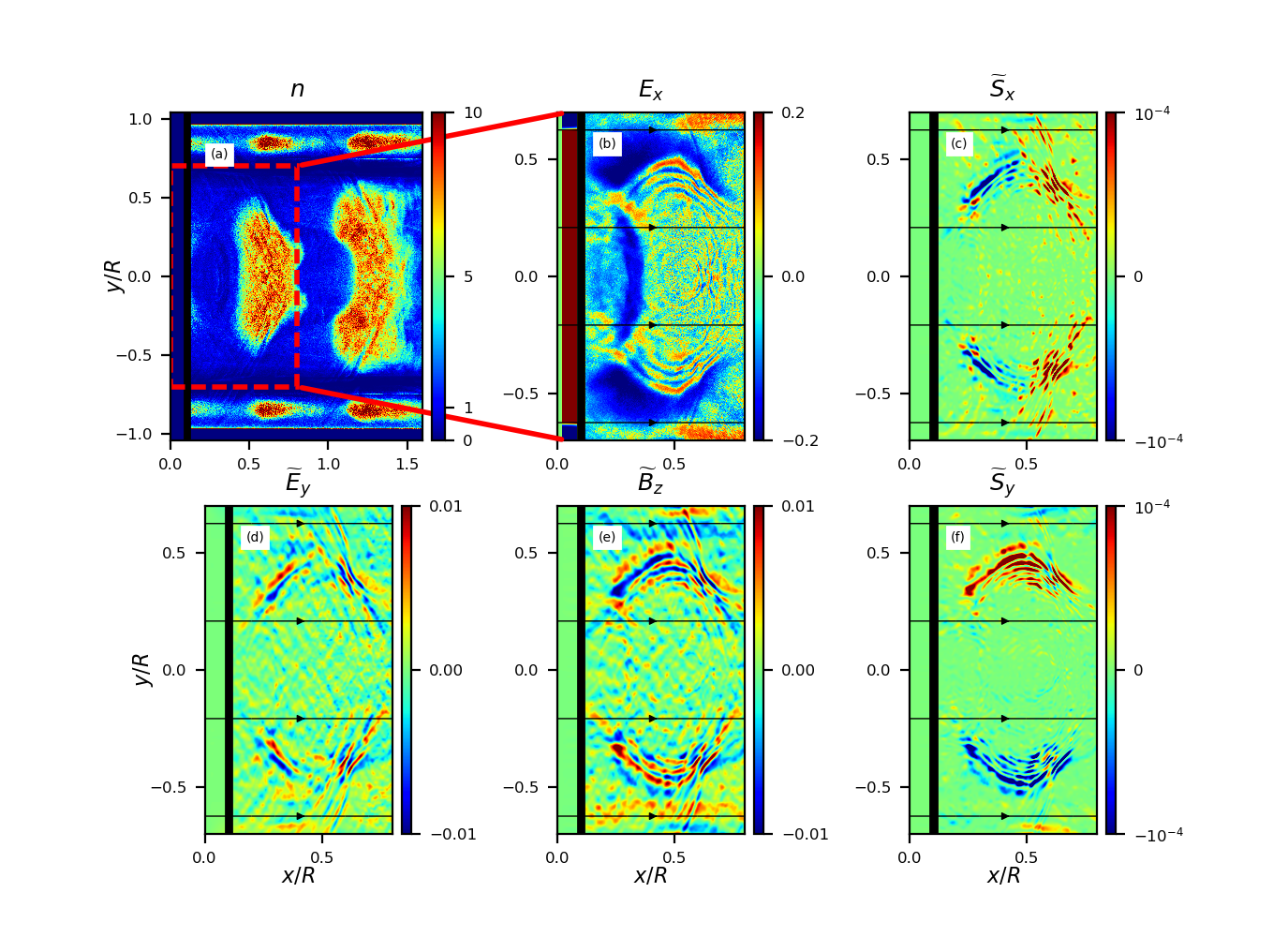}
   \caption{Formation of the electromagnetic wave during the process of dynamic screening of the electric field in the pair plasma discharge. Individual panels show two-dimensional distributions of (a): plasma density ($n$), normalized to the minimum density required to carry  the current, $V_0 B_0/4e\pi c R$; (b) electric field component along the background magnetic field ($E_x$); electromagnetic field quantities computed by subtracting the $x$-averaged distributions to better visualize the wave pattern: (d) transverse component of the electric field $\widetilde{E}_y$; (c) longitudinal component of the Poynting flux vector $\widetilde{S}_x$; (e) out-of-plane component of the magnetic field $\widetilde{B}_z$; (f) transverse component of the Poynting flux vector $\widetilde{S}_y$. The electromagnetic field components are normalized to vacuum field, $(V_0/c)B_0$, and Poynting flux components are normalized to the bulk Poynting flux, $S_0=c(V_0/c)^2 B^2_0 /4\pi$, that a sheared conductor launches. In all panels distances are measured in units of the conductor's half-length, $R$.}
    \label{fig:discharge}

\end{figure*}
\begin{figure*}
    %\label{fig:my_label}
        \includegraphics[width=1.\textwidth, trim = 50.0mm 20.0mm 50.0mm 5.0mm]{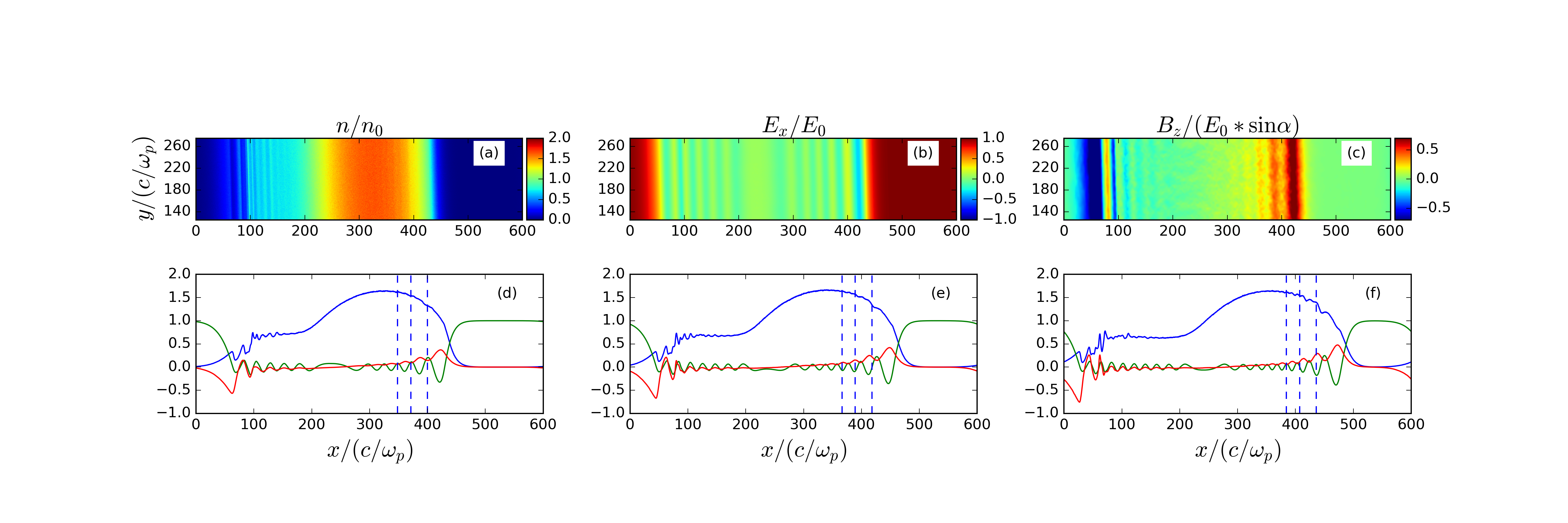}
   \caption{Formation of the superluminal electromagnetic wave during the process of dynamic screening of the electric field. First row shows two-dimensional distributions of (a): plasma density ($n$); (b) electric field component along the background magnetic field ($E_x$); (c) out-of-plane component of the magnetic field ($B_z$), at $0.47L_x/c$, where $L_x$ is the size of the computational box in the $x$-direction. Panels (d)-(f) show the snapshots of $n$ (blue line), $E_x$ (green line) and $B_z$ (red line) at three consecutive moments of time in the simulation, $0.47L_x/c$, $0.53L_x/c$ and $0.6L_x/c$. Blue vertical dashed lines move to the right at the speed of light. The wavefront clearly overruns the position of the lines, which proves the superluminal phase speed of the mode. Panels (a)-(c) show zoom-in onto the central part of our numerical domain which is larger in the $y$-direction. Simulation is performed  for angle between the normal to plasma injection front and the background magnetic field $\alpha=1/40$. In all panels distances are measured in units of the plasma skin depth, $c/\omega_p$, where plasma frequency $\omega_p=\sqrt{4\pi e^2 n_0/m_e}$ is calculated for the cold plasma of a fiducial density $n_0$.}
    \label{fig:screening}

\end{figure*}

We simulate pair plasma discharge in a simplified setup in a Cartesian two-dimensional computational box. We use a relativistic particle-in-cell code TRISTAN-MP \cite{Spitkovsky05}. On the left side of the box, we put a conducting plate which represents a NS, and immerse it into a background magnetic field, $B_0$, which is set to be along the $x$-axis. Inside the conductor, the electric field is forced to corotation values, ${E_y(y)=-(V_0-{\beta}_0) B_0(y-y_0)g(y)/c R}$, where $y$ is the direction perpendicular to the background magnetic field, $y_0$ is the position of the conductor’s center, $R$ is its half-length, $g(y)$ is the smoothing function that drives the electric field to zero at the edge of the conductor, $V_0$ is the amplitude of the linear velocity of rotation, and $\beta_0$ describes the general-relativistic inertial frame-dragging effect (see Supplemental material). The normal component of the magnetic field, $B_x=B_0$, is fixed at the conductor's surface. To mimic the surface charge extraction from the NS atmosphere, we inject neutral pair plasma at the boundary of the conductor. Once an electron or positron reaches the threshold energy, $\gamma_{\rm th} m_e c^2$, it begins to emit curvature photons capable of pair production. Emission and propagation of photons is done with a Monte Carlo technique, and the mean free path of photons is $R/5$. Momentum of emitted high-energy photons is directly removed from particle's momentum. To mimic spatially varying curvature of the magnetic field that determines the photon energies, we set the threshold for pair production to depend on the transverse coordinate: $\gamma_{\rm th}(y)=\gamma_{\rm 0}(1+(y/R)^2)$. The simulation presented below is carried out for the following values of numerical parameters: $V_0=2\beta_0=0.2 c$, $\gamma_{\rm 0}=5\times 10^{-3}\gamma_{\rm tot}=100$, where $\gamma_{\rm tot} = e B_0 (V_0/c) (R/2) / m_e c^2 $ is the Lorentz factor of a particle experiencing the full vacuum potential drop across the conducting plate. A plasma skin depth calculated  for  the  cold  plasma  of a Goldreich-Julian (Ref. \cite{Goldreich69}; hereafer, GJ) density (the minimal density of plasma needed to screen the accelerating electric field in the magnetosphere of a rotating magnetized NS), ${n_{\rm GJ}} = V_0 B_0/4\pi c R e$, is resolved with $30$ numerical cells. Though our simplified setup does not model actual physical conditions in the pulsar polar cap, the choice of numerical parameters in our experiment captures the physics of particle acceleration, pair production and dynamics of the electromagnetic fields in the polar cap accelerator (see Supplemental material).

Simulation starts in vacuum, and after few light crossing times along the conductor, $R/c$, it reaches a quasi-steady state of repeating pair plasma production episodes followed by quiet states. In Fig.~\ref{fig:discharge}a we show a representative snapshot of plasma density produced in the simulation, which shows two pair plasma clouds produced in subsequent discharge episodes. The gap
%, or the region of non-zero parallel electric field where significant particle acceleration occurs, 
appears close to the surface of the plate. Previous episode of pair formation resulted in the plasma cloud located at $x\approx 1.2$, and we focus on the cloud located at $x\approx0.5$. The screening of the accelerating electric field proceeds in the form of parallel electric field fluctuations clearly visible in Fig.~\ref{fig:discharge}b. The non-uniformity of pair formation in the transverse ($y$-) direction results in inhomogeneity of the electric field across magnetic field lines, i.e., in a non-zero $\nabla_y \times E_x$. This directly leads to the production of an out-of-plane fluctuating component of the magnetic field $B_z$. The inhomogeneity of the pair formation along the magnetic field leads to a non-uniform $B_z$, i.e., to a non-zero $\nabla_x \times B_z$, which gives rise to a wave component $E_y$ (the details of the mechanism generating transverse fields are presented in Supplemental material). The presence of transverse fluctuating components $\widetilde{E}_y$ and $\widetilde{B}_z$, clearly visible in Fig.~\ref{fig:discharge}(d),\ref{fig:discharge}(e) in regions where the pair formation front is inclined to the magnetic field, strongly suggests the electromagnetic nature of the wave. Here, $\widetilde{B}_z$ and $\widetilde{E}_y$ are computed as $\widetilde{B}_z=B_z-\langle B_z \rangle_x$, where $\langle\rangle_x$ represents spatial averaging along the $x$-direction and is performed to subtract the zero-frequency out-of-plane component of the magnetic field created by the bulk plasma current. The waves are linearly polarized, and their transverse electric field vector lies in $\vec{k}$-$\vec{B}$ plane (here, $\vec{k}$ is the wavevector of propagating waves, directed at a non-zero angle with respect to the external magnetic field). This suggests that an ordinary mode is excited in the process of screening the electric field. Electric field fluctuations are accompanied by a significant transverse magnetic field component and Poynting flux only if pair production front is inclined to the local magnetic field. For example, in panels (c) and (f) of Fig.~\ref{fig:discharge} the wave Poynting flux, computed as $\widetilde{{\bf S}}={\bf S}-\langle {\bf S} \rangle_x$, is nearly zero in the middle of the pair cloud, $y\approx 0$, where screening happens almost perpendicular to the background magnetic field direction. However, at $y/R\approx \pm 0.5$, where the screening happens at a non-zero angle to the magnetic field, the wave flux reaches $\approx 10^{-4} S_0$, where $S_0=c (V_0/c)^2 B^2_0/4\pi$ is the bulk Poynting flux that a sheared plate launches. Interestingly, comparable wave power is emitted in both forward and backward directions. The waves that are emitted towards the NS are later reflected from the stellar surface and also propagate outwards. The discharge and wave emission repeat after most of the freshly produced pair plasma escapes from the gap region, and the accelerating electric field is restored. The timescale of this variability is slightly longer than the gap's light-crossing time.

\begin{figure*}
    \label{fig:my_label}
        \includegraphics[width=1.\textwidth]{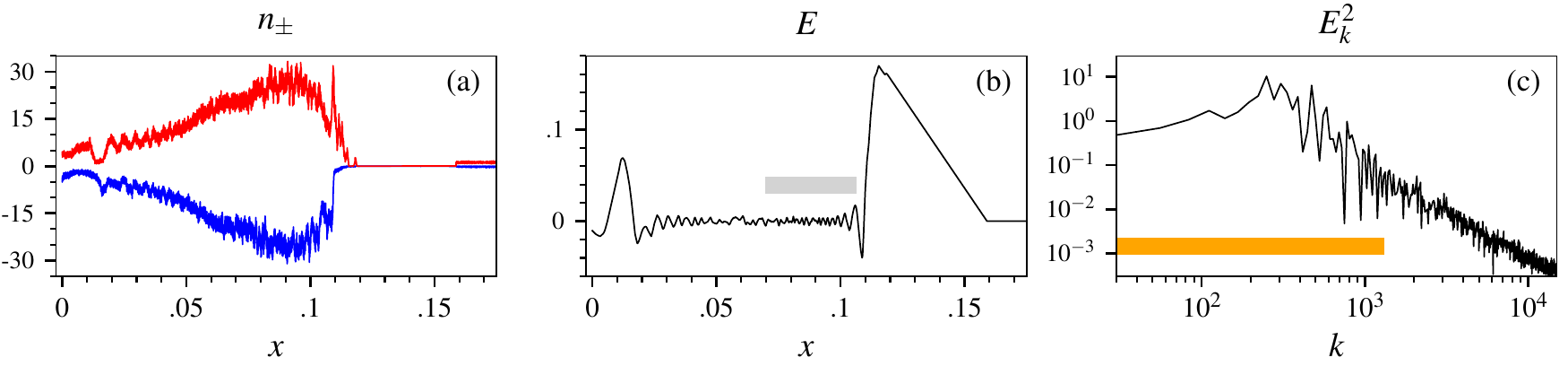}
   \caption{One-dimensional pair discharge for realistic parameters ($P=33$~ms, $B=10^{12}$~G, magnetic field line radius of curvature $\rho_c=1.67\cdot10^{7}$~cm). Shown are: (a): charge densities $n_{\pm}$ of positrons (red line) and electrons (blue line), normalized to $n_{{\rm GJ}}$, (b):  electric field $E$ normalized to the vacuum field $E_{vac}=\Omega{}B r_{pc}/2c=7.6\cdot10^{10}$~V/cm, and (c): power spectrum of the electric field $E_k^2$ for spatial intervals shown by the gray bar in panel (b). The orange bar in panel (c) indicates the range of spatial frequencies  $k$, measured in units of $1/L$, where $L=0.25 r_{pc}$ is the size of the simulation box, corresponding to the range of  plasma skin depths in the region shown by the gray bar in panel (b). Distance $x$ from the NS is normalized to the polar cap radius $r_{pc}=7.97\cdot10^{4}$~cm.}
    \label{fig:power_spectrum}
\end{figure*}

To better understand the properties of these electromagnetic waves, we perform local 2D simulations of electric field screening during the burst of pair formation (as observed in our simulations described above) in a controlled setup, which can be considered as a zoom-in on pair formation fronts seen in Fig.~\ref{fig:discharge}. In this experiment, we start with a constant electric field, $E_0$, along the background magnetic field, both pointed along the $x$-axis. We then inject pair plasma at a constant rate $\dot{n}$ in a slab (see Fig.~\ref{fig:screening}a), which moves at $0.999c$ along the $x$-axis. The normal of the slab is inclined to the background magnetic field by angle $\alpha$, which we vary. Pairs are injected moving with Lorentz factor $\gamma=4$ along the direction of magnetic field. The screening of the electric field happens in two stages. Initially, freshly injected electrons and positrons are quickly accelerated by the vacuum electric field in the opposite directions, which creates a strong current $\approx 2n_e e c=2\dot{n} e c t$, where $n_e$ is the local electron density. This current increases as the density of pair plasma increases, and leads to the quick screening of the electric field in time $\tau \approx \sqrt{E_0/(4\pi \dot{n} e)}$. At some point the electric field gets sufficiently low and cannot reverse the direction of motion of freshly injected pairs. After this happens, the electric field starts to oscillate, with an amplitude that slowly decreases as the plasma density increases. As more charges are available to screen the electric field, less charge separation is required for screening, and the wavelength of the oscillation decreases. The snapshots of parallel electric field component, $E_x$, and of transverse magnetic field component, $B_z$, are shown in panels (b) and (c) of Fig.~\ref{fig:screening}, respectively. By performing simulations for different values of the angle $\alpha$, we confirm that the amplitude of the out-of-plane  component of the magnetic field in the wave scales as $B_z\approx(k_{\perp}/k_{\parallel}) E_x =E_x \sin\alpha$. In panels (d)-(f) of Fig.~\ref{fig:screening} we show horizontal slices of plasma density, electric and magnetic field components along the center of the box. We plot three vertical lines that move with the speed of light in the direction of the injected plasma. The wavefront clearly overruns the position of the lines, which suggests that the phase speed of the wave exceeds the speed of light, i.e., the excited mode is superluminal. These two facts, polarization and superluminal character, lead us to the conclusion that the wave that gets excited in the process of the discharge is an electromagnetic O-mode of highly magnetized pair plasma \cite{Arons86}. We also find that the shape of the time series of the out-of-plane component of the magnetic field in the wave is essentially controlled by the behavior of the parallel electric field, or, by the dynamics of the discharge.

To test that the proposed mechanism can generate plasma waves with wavelengths comparable with those of pulsar radio emission we perform high resolution discharge simulations for unscaled physical parameters. We use the 1D hybrid PIC-Monte-Carlo code PAMINA \cite{Timokhin10,Timokhin13} which models polar cap discharges for realistic values of all physical parameters (see Supplemental material for details of the numerical setup). A snapshot from these simulations is shown in Fig.~\ref{fig:power_spectrum}, with values of physical parameters listed in the caption. We show the charge densities of positrons and electrons $n_\pm$, the electric field $E$, and the power spectrum of the electric field $E_k^2$ for the region where the outgoing wave is formed, shown by the gray bar in the plot for $E$.  The emerging wave occupies a broad range of spatial frequencies visible as large amplitude fluctuations superimposed on the power-law-like part of the spectrum.  Both spatial and energy distributions of newly created pair plasma are highly inhomogeneous, and the plasma frequency $\omega_p=\sqrt{4\pi e^2 \langle n/\gamma^3 \rangle /m_e}$ varies by a few orders of magnitude. The range of $k$ corresponding to the range of skin depths, $\lambda_D=c/\omega_p$, in the region where the wave is forming is shown with the orange bar. The wave spectrum already spans an order of magnitude, with the shortest wavelengths reaching $\lambda_D$ (not surprisingly, as the latter is the characteristic scale for the screening of the electric field). As the pair formation continues, $\lambda_D$ decreases and the range of wavelengths of the plasma wave increases. Hence, the emergent coherent radiation, when it decouples from the plasma, will be broadband, extending up to the plasma frequency of pair plasma at the point of decoupling. The superluminal wave is generated by continuous pair injection, permitting the screening of the electric field on shorter and shorter scales. When the pair formation stops, so should the wave generation.  The highest frequency of coherent radiation produced by this mechanism can be estimated as  $\nu\simeq\sqrt{4\pi e^2 \kappa{n_{\rm GJ}}/\langle \gamma^3 \rangle m_e}/2\pi=26\sqrt{\kappa_5 B_{12}/r^{3}P_{0.1}\gamma_{10}^3}$~GHz, where $\kappa_{5}$ is the final pair multiplicity normalized to $10^5$, $B_{12}=B/10^{12}$G, $P_{0.1}=P/0.1$s,  $r$ -- distance from the NS in NS radii, and $\gamma_{10}=\gamma/10$ is the Lorentz factor of final pair generation. The highest pair multiplicity is usually achieved at the last cascade generation, at distances from the NS comparable to NS radius \cite{Timokhin19}.

To summarize, by means of first-principles kinetic plasma simulations we presented a robust pulsar radio emission mechanism -- this mechanism does not require any special physical conditions in pulsar polar caps, it should always work if pair discharges are intermittent and non-uniform across the magnetic field. These two conditions are expected to be fulfilled in polar caps of all pulsars (e.g., Refs. \cite{Arons83,Timokhin13}). We found that if the pair formation front is inclined to the background magnetic field, the non-steady plasma discharge produces broad-band emission in the form of the electromagnetic O-mode with a spectrum controlled by the dynamics of the discharge. In pulsars, the pair formation front is inclined to the magnetic field due to, e.g., transverse variation of magnetic field line curvature or the variation of the accelerating electric field. This makes the regions near polar cap edges (where the accelerating electric field sharply jumps at the return current layer) and near magnetic poles (where the radius of curvature of magnetic field lines rapidly approaches infinity) the most efficient generators of radio emission, which might explain the existence of cone and core components in pulsar radio profiles \cite{Rankin90}.

This mechanism significantly differs from the usually postulated ad hoc models that require either the conversion of plasma modes from the two-stream instability or the curvature radiation from charged bunches. In fact, neither 1D nor 2D simulations of non-stationary discharge show the  formation of charge density clumps or signs of streaming instabilities. It is the intrinsically time-dependent nature of plasma discharge that drives coherent radio emission \cite{Beloborodov08,Timokhin10,Timokhin13}. While we find robust excitation of O-modes, pulsar radio polarimetry observations suggest the presence of both O and X-modes. We suggest the following mechanism that can contribute to the production of the X-mode. At some distance from the star, the O-mode produced in the discharge process can escape from the dense plasma cloud into a rarefied zone, where its polarization characteristics freeze. As it propagates away, the wave may encounter a dense plasma with a local magnetic field different from that where it was emitted. 
%This suggests that 
There the mode is no longer an eigenmode of the plasma, and may experience conversion into other plasma modes, in particular into an X-mode.  
%We will investigate this possibility with future simulations. 

\begin{acknowledgments}
The authors would like to thank J.\ Arons, A.\ Beloborodov, V.\ Beskin, A.\ Jessner, and Yu.\ Lyubarsky for numerous insightful discussions. AP and AS acknowledge hospitality of KITP, where part of this work was performed. This work was supported by NASA grants NNX15AM30G and 80NSSC18K1099 and by the National Science Foundation under Grants No. NSF AST-1616632 and NSF PHY-1748958. Resources supporting this work were provided by the NASA High-End Computing Program through the NASA Advanced Supercomputing Division at Ames Research Center. Research at the Flatiron Institute is supported by the Simons Foundation, which also supported AS (grant 267233). We also thank the anonymous referees for valuable comments on the Letter.
\end{acknowledgments}

%\bibliography{bibfile.bib}

%\end{document}
%\bibliography{bibfile.bib}

%%%%%%%%%% Merge with supplemental materials %%%%%%%%%%
\pagebreak
\widetext
\begin{center}
\textbf{\large Supplemental Materials}
\end{center}
%%%%%%%%%% Merge with supplemental materials %%%%%%%%%%
%%%%%%%%%% Prefix a "S" to all equations, figures, tables and reset the counter %%%%%%%%%%
%\setcounter{equation}{0}
\setcounter{figure}{0}
\setcounter{table}{0}
\makeatletter
\renewcommand{\theequation}{S\arabic{equation}}
\renewcommand{\thefigure}{S\arabic{figure}}
%\renewcommand{\bibnumfmt}[1]{[S#1]}
%\renewcommand{\citenumfont}[1]{S#1}
%%%%%%%%%% Prefix a "S" to all equations, figures, tables and reset the counter %%%%%%%%%%

In this supplemental material we provide additional details about waves in pulsar magnetosphere and our numerical setups.

\section{Overview of waves in Pulsar Plasma}
Strongly magnetized pair plasma near the pulsar surface supports a very limited set of eigenmodes because charged particles can only move along the magnetic field lines. Assuming the plasma is cold, the wave dispersion relation in the limit of infinite magnetic field strength is \cite{Arons86,Beskin93}: $(\omega^2-c^2k^2_{\parallel})(1-\omega^2_p/\omega^2)-c^2k^2_{\perp}=0$, where $\omega_p$ is the plasma frequency, $\omega$ is the frequency of the wave, and $k_{\parallel}$ and $k_{\perp}$ are the wave vector components parallel and perpendicular to the background magnetic field, respectively. Here, all quantities are calculated in the rest frame of the plasma that streams relativistically at Lorentz factor $\gamma$. This equation describes three waves: extraordinary mode, with electric field vector perpendicular to both the wave vector and the background magnetic field, and two modes on the ordinary branch: the superluminal mode and the subluminal Alfven mode. The extraordinary mode does not interact with the plasma, which means it is hard to excite this mode with plasma currents. The Alfven mode propagates along magnetic field lines and suffers from Landau damping \cite{Arons86}. The superluminal O-mode, on the other hand, satisfies a lot of the requirements for explaining pulsar radio emission. First, since the mode is superluminal, it is not damped by Landau mechanism, so it can freely propagate in the pulsar plasma. Second, the mode is partially electromagnetic if it propagates at a non-zero angle with respect to the background magnetic field. For small angles of propagation, $k_{\perp}/k_{\parallel}=\sin\alpha$, the ratio of electromagnetic and electrostatic field components in the wave satisfies $B_{\perp}/E_{\parallel} \propto \sin \alpha$. Third, as this mode propagates in a plasma of decreasing density, it becomes a freely propagating electromagnetic mode. Thermal effects lead to quantitative corrections to the wave dispersion, but do not change the qualitative behavior of the three modes \cite{Melrose19,Beskin19}.

\section{Generation of transverse waves in non-uniform discharges}

Consider a discharge in a super strong curl-free background magnetic field directed along $x$-axis. Particles can move only along the magnetic field lines and the only non-zero component of electric current is $j_x$. Let us assume that the system is uniform in $z$-direction, so that derivatives $\partial_z$ of all physical quantities are zero. In this case, Maxwell equations for electromagnetic field components can be divided into two sets of equations. The first set
\begin{eqnarray}
    \partial_x B_y - \partial_y B_x & = & \frac{1}{c}\,\partial_t E_z 
    \label{Am_Ez},\\
    \partial_y E_z & = & - \frac{1}{c}\,\partial_t B_x 
    \label{Far_Bx},\\
    \partial_x E_z & = &   \frac{1}{c}\,\partial_t B_y,
    \label{Far_By}
\end{eqnarray}
describes the evolution of a mode with components $E_z,B_x,B_y$. Equations~(\ref{Am_Ez})-(\ref{Far_By}) do not contain the current term and this (extraordinary) mode does not couple to the plasma. Hence, it cannot be directly emitted in the discharge. The second set of equations 
\begin{eqnarray}
    \partial_y B_z & = & \frac{4\pi}{c}j_x + \frac{1}{c}\,\partial_t E_x \label{Amp_Ex},\\
    -\partial_x B_z & = & \frac{1}{c}\,\partial_t E_y 
    \label{Amp_Ey},\\
    \partial_x E_y - \partial_y E_x & = & - \frac{1}{c}\,\partial_t B_z,
    \label{Far_Bz}
\end{eqnarray}
describes a mode with components $E_x,E_y,B_z$. This mode can couple to the plasma because of the current term $j_x$ in eq.~\ref{Amp_Ex}. However, a transverse mode can only be excited if the system is non-uniform in the  $y$-direction.  If the discharge is uniform in $y$, the derivatives $\partial_y$ are zero, and $E_x$ decouples from $E_y$ and $B_z$. In this case the evolution of $E_x$ coupled to $j_x$ is completely determined by eq.~\ref{Amp_Ex}, and evolution of $E_y,B_z$ by eqs.~\ref{Amp_Ey},~\ref{Far_Bz} which are independent of eq.~\ref{Amp_Ex}. Therefore, the non-uniformity of the discharge across magnetic field lines is a necessary condition for excitation of transverse waves. 

The qualitative picture of how transverse waves are excited in discharges is as follows. Injection of particles due to pair formation gives rise to the electric current $j_x$  driven by accelerating electric field of the gap. $j_x$ induces fluctuating electric field $E_x$ (eq.~\ref{Amp_Ex}). The non-uniformity of $E_x$ across magnetic field lines gives rise to the fluctuating magnetic field $B_z$ (eq.~\ref{Far_Bz}). The non-uniformity of the whole process along magnetic filed lines (screening proceeds on smaller and smaller scales) induces fluctuating perpendicular electric field (eq.~\ref{Amp_Ey}), thus coupling all field components.

\section{Guiding center particle pusher}
In order to accurately represent particle motion in the strong magnetic field of pulsars, we use a new particle pusher algorithm that solves only for the motion of the guiding center of the orbit and, thus, completely eliminates gyrational motions of the particles. This is the appropriate physical regime for field strengths near pulsars, where the typical synchrotron loss time for secondary pairs in the first Landau level, $10^{-14} {{\rm s}}$, is much shorter than any other characteristic timescale.

We solve the guiding-center equations, ${\rm d}{\vec{x}}/{\rm d}t = {\vec{v}}(\vec{x})= v_{\parallel} {\vec{B}}/B + c {\vec{E}_{\perp}}\times {\vec{B}}/B^2$ and ${\rm d} \vec{p}_{\parallel}/{\rm d}t = e \vec{E}_{\parallel}$. Here, $\vec{x}$ is particle's coordinate, $\vec{p}_{\parallel}=m_e \gamma {\vec{v}}_{\parallel}$ and ${\vec{v}}_{\parallel}$ are particle's momentum and velocity along the magnetic field, $\gamma=1/\sqrt{1-v^2_{\parallel}/c^2-E^2_{\perp}/B^2}$ is the particle's full Lorentz factor, and $E_{\parallel}$ and $E_{\perp}$ are electric field components along and perpendicular to the local magnetic field, respectively. Here we neglect all drifts associated with field curvature and time-dependence, as is appropriate for high magnetic field strengths of pulsars. We solve guiding-center equations using a modified leapfrog algorithm. On every timestep we update the particle's momentum along the magnetic field as ${\vec p}_{\parallel, \rm new}={\vec p}_{\parallel, \rm old}+eE_{\parallel}(\vec{x}_{{\rm old}})$, and then update particle's coordinate as ${\vec x}_{{\rm new}}-\vec{x}_{{\rm old}}=0.5(\vec{v}(\vec{x}_{{\rm old}})+\vec{v}(\vec{x}_{{\rm new}})) \Delta t $, where $\Delta t$ is the timestep of the code, using the fixed point iteration method.

\section{2D discharge simulations}

{\bf{General-relativistic frame-dragging.}} In flat space-time, $\beta_0=0$, and assuming the presence of dense pair plasma, our numerical setup drives an outflow with a GJ current, $J_x= - V_0 B_0/4\pi R= {\rho_{\rm GJ}} c$, ${\rho_{\rm GJ}} = - V_0 B_0/4\pi c R$, and represents the plasma conditions near the axis of an aligned pulsar \cite{Timokhin06}. As was shown by \cite{Philippov15}, general-relativistic corrections make the current to be super-GJ, which triggers efficient pair production \cite{Mestel85,Beloborodov08,Timokhin13}. We model general-relativistic effects by adding a term into Faraday’s induction equation, which describes the generation of an electric field due to the rotation of spacetime \cite{Philippov18}: ${\partial{\vec B}}/{\partial t}/c=-\nabla \times \left(\vec{E} +\vec{\beta}/c\times \vec{B}\right)$, where ${\beta}_y = {\beta_0} (y-y_0) (2R/(x-x_0))^3/R$ is the only non-zero component of $\vec{\beta}$, and $x_0$ is the position of the conductor's center. In the steady-state, the current, $J_x$, is the same as in the flat space-time setup, but the local charge density near the star is reduced \cite{Beskin90}, $\rho \approx - (V_0-\beta_0) B_0/4\pi c R$. This ensures the current flow to be super-GJ, which leads to a non-stationary discharge and efficient production of electron-positron pairs \cite{Mestel85,Beloborodov08,Timokhin13}. Our particular choice $V_0=2\beta_0$ leads to a super-GJ current $J_x\approx 2\rho c$, which is typical for the magnetosphere of the oblique pulsar \cite{Timokhin13,Philippov18}.

{\bf{Surface charge injection.}} At the boundary of the conductor we inject neutral pair plasma at a rate $(0.2E_{\parallel})/ 4\pi e $ per time step, where $E_{\parallel}=\lvert{\bf E}\cdot {\bf B}\rvert/{\rm B}$ is the component of the electric field along the magnetic field at the injection point. This prescription mimics free supply of particles in the NS atmosphere. For lower injection rates we find that electric field at the plate is not sufficiently screened (this case corresponds to the models with no particle extraction from the surface \cite{Ruderman75}), while larger rates lead to virtual cathode oscillations. Even though we inject neutral plasma, the particles are injected at rest, and one sign of charge is pulled into the plate, while the other is accelerated outwards. On the other hand, it has been shown that discharges with both free escape of particles and no escape of particles operate in a similar intermittent fashion \cite{Timokhin10,Timokhin13}, so our conclusions on the generation of radio waves should  be insensitive to this prescription.  

{\bf{The choice of numerical parameters}}. For a typical young energetic pulsar the full vacuum potential drop across the polar cap corresponds to particle Lorentz factor  $\gamma_{\rm tot}\approx (\Omega R/c)^2(eB R/m_e c^2)\sim10^{10}$, where $\Omega$ is the rotational frequency of the NS, $R$ is the NS radius, and $B$ is the magnetic field strength near the surface. Energetic electrons and positrons emit pair-producing curvature photons once they reach energies $\gamma_{\rm th}m_e c^2$, where $\gamma_{\rm th}\sim 10^6$. The typical Lorentz factor of secondary pairs produced in a process of the pair discharge near the stellar surface is in the range $\gamma_{\rm s} \sim 10^2-10^3$. The multiplicity of the produced pair plasma reaches $\sim 10^5$ in the last cascade generation. It is not possible to conduct a direct multi-dimensional PIC simulation for such high values of particle Lorentz factors and plasma densities. In our simulation we scale down these numbers, $\gamma_{\rm tot}\sim {10^4}$, $\gamma_{\rm th}\sim {10^2}$ and $\gamma_{\rm s}\sim {\rm few}$, but preserve the hierarchy $\gamma_{\rm s}\ll \gamma_{\rm th} \ll \gamma_{\rm tot}$. This choice of parameters allows copious production of electron-positron pairs and time-dependent behavior of the discharge which is similar to 1D PIC simulations performed for realistic parameters \cite{Timokhin13}. The transverse dependence of the pair production threshold $\gamma_{\rm th}(y)\sim1+(y/R)^2$ is chosen to represent the simplest non-homogeneity across magnetic field lines. The mean free-path of energetic curvature photons is comparable to the size of the particle acceleration zone \cite{Ruderman75}, which is preserved in our simulations.

\section{1D discharge simulations}
\begin{figure*}
    %\label{fig:my_label}
    
        \includegraphics[width=1.\textwidth]{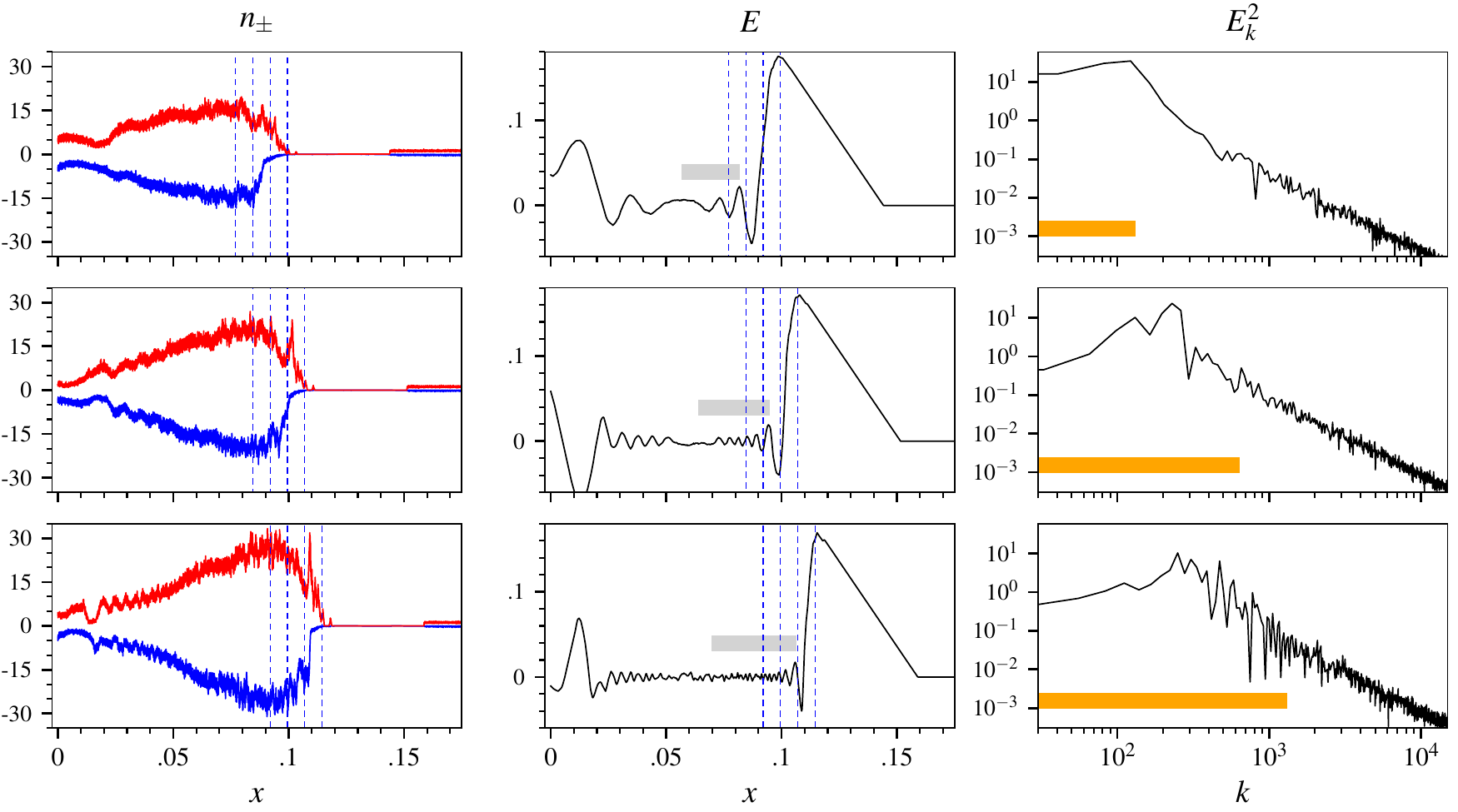}
        \caption{One-dimensional pair discharge in polar cap for realistic
          parameters ($P=33$~ms, $B=10^{12}$~G, magnetic field line radius of
          curvature $\rho_c=1.67\cdot10^{7}$~cm). Shown are three snapshots at
          different stages of electric field screening for (left panels): charge
          densities $n_{\pm}$ of positrons (red line) and electrons (blue line),
          normalized to $n_{\rm GJ}$; (middle panels): electric field $E$ normalized
          to the vacuum field
          $E_{vac}=\Omega B r_{pc}/2c=7.6\cdot10^{10}$~V/cm, and (right
          panels): power spectrum of the electric field $E_k^2$ for spatial
          intervals shown by grey bars on the middle panels. Orange bars in the
          right panels indicate the range of spatial frequencies  $k$, measured in units of $1/L$, where $L=0.25 r_{pc}$ is the size of the simulation box, corresponding to the range of plasma skin depths in the regions shown by the gray bars. Distance $x$ from the NS is normalized to the polar cap radius
          $r_{pc}=7.97\cdot10^{4}$~cm. Blue vertical dashed lines move to the
          right at the speed of light. The earliest snapshot is in the first row, the latest one -- in the third (it is the same snapshot as shown in Fig. 3 in the paper).}
    \label{fig:suppl}

\end{figure*}

It has been shown that qualitative cascade behavior does not depend on the boundary conditions on NS surface \cite{Timokhin13}, and so we can resort to the simplest case to check the scaling of the emergent emission. We model discharges for the pulsar model with no particle extraction from the NS surface \cite{Ruderman75}; this allows us to model polar cap discharges without invoking scaling of physical parameters. We model discharges along a single magnetic field line, the magnetic field is anti-parallel to the angular velocity of NS rotation, and the GJ charge density ${\rho}_{\rm GJ}$ is positive; the imposed current density through the domain is $j={{\rho}_{\rm GJ}}c$. Simulations are performed with the hybrid PIC/Monte-Carlo code PAMINA \cite{Timokhin10,Timokhin13} which self-consistently models particle acceleration and evolution of the electric field via PIC algorithm, and high energy photon emission, propagation, and pair creation via Monte-Carlo algorithm for realistic physical parameters. We model discharges along a dipolar magnetic field line with the radius of curvature $\rho_c=1.67\times 10^{7}$~cm in the polar cap of a pulsar with the period $P=33$~ms and magnetic field strength $B=10^{12}$~G. We start the simulations with numerical domain filled with dense plasma having a GJ charge density, and then let it evolve. As described in \cite{Timokhin10}, after a short relaxation time, the system settles down to a limit cycle behavior with quasi-periodic bursts of pair formation. Particles in this simulation reach energies up to $\gamma\sim10^8$ and emit curvature radiation photons which are absorbed in the strong magnetic field and create electrons and positrons via single photon pair production.

In Fig.~3 in the paper we show a single snapshot from this simulation. Here we want to illustrate the development of the wave in more detail. The three snapshots in Fig.~\ref{fig:suppl} show the temporal evolution of the wave. The earliest snapshot is in the first row, the latest -- in the third one, which is the same snapshot as shown in Fig.~3 of the paper. We plot the charge densities of positrons and electrons
$n_\pm$, the electric field $E$, and the power spectrum of the electric field
$E_k^2$ for the regions where the outgoing wave is formed, shown by gray bars in
the panels that show $E$.  We also plot three blue vertical dashed lines which move to the right at the speed of light.  It is easy to see that the wave is superluminal as the peaks in the distribution of the electric field $E$ overrun these lines.  In the first snapshot, the emerging wave results in a broad
peak in the power spectrum; in the second and third snapshots, it occupies a broader range of spatial frequencies. In these power spectra plots the wave corresponds to large amplitude fluctuations superimposed on the power-law-like parts of the spectra (the size of the numerical grid cell corresponds to $k=40000$, so these fluctuations are well resolved). With time the wave extends over larger spatial region (shown by gray bars).  The range of $k$ which corresponds to the range of plasma skin depth in the gray region is shown with the orange bar. The shortest wavelengths in the wave power spectrum are comparable with the plasma skin depth $\lambda_D$. As the pair formation continues, $\lambda_D$ decreases and the spectrum of the wave, which coincides with the orange region in spectral plot, extends to higher frequencies.

\end{document}